\documentclass[preprint,showpacs,amssymb,amsmath,aps,pra]{revtex4}

\usepackage{graphics}
\usepackage{bm}
\begin{document}
\title{Entanglement generation resonances in XY chains}
\author{R. Rossignoli, C.T. Schmiegelow}
\affiliation{Departamento de F\'{\i}sica-IFLP,
Universidad Nacional de La Plata, C.C.67, La Plata (1900), Argentina}
\begin{abstract}
We examine the maximum entanglement reached by an initially fully aligned state
evolving in an $XY$ Heisenberg spin chain placed in a uniform transverse
magnetic field. Both the global entanglement between one qubit and the rest of
the chain and the pairwise entanglement between adjacent qubits is analyzed. It
is shown that in both cases the maximum is not a monotonous decreasing function
of the aligning field, exhibiting instead a resonant behavior for low
anisotropies, with pronounced peaks (a total of $[n/2]$ peaks in the global
entanglement for an $n$-spin chain), whose width is proportional to the
anisotropy and whose height remains finite in the limit of small anisotropy. It
is also seen that the maximum pairwise entanglement is not a smooth function of
the field even in small finite chains, where it may exhibit narrow peaks above
strict plateaus. Explicit analytical results for small chains, as well as
general exact results for finite $n$-spin chains obtained through the
Jordan-Wigner mapping, are discussed.
\end{abstract}
\maketitle

\section{Introduction}
Quantum entanglement has been long recognized as one of the most fundamental
and intriguing features of quantum mechanics \cite{S.35}. It denotes the
ability of composite quantum systems to develop correlations which have no
classical counterpart. Interest on all aspects of entanglement has grown
enormously since its potential for permitting radically new forms of
information transmission and processing was unveiled
\cite{Be.93,Ek.91,Di.95,Be.00}, being now considered an essential {\it
resource} in the field of quantum information science \cite{NC.00}, where
rigorous entanglement measures have been introduced \cite{Be.96,W.98}. The
interest has also extended to other areas like condensed matter physics, where
it has provided a novel perspective for the analysis of correlations and
quantum phase transitions \cite{ON.02,OS.02,V.03,VW.04,T.04}.

Spin systems with Heisenberg interactions \cite{LSM.61,S.99} constitute a
particularly attractive scenario for studying quantum entanglement. They
provide a scalable qubit representation suitable for quantum processing tasks
\cite{LDV.98,I.99,Bos.04,L.02} and can be realized by diverse physical systems
such as cold atoms in optical lattices \cite{D.03}, quantum dots
\cite{LDV.98,I.99} and Josephson junctions arrays \cite{MSS.01}. Accordingly,
several investigations  of  entanglement in ground and thermal equilibrium
states of Heisenberg spin chains subject to an external magnetic field have
been made (see for instance \cite{ON.02,OS.02,V.03,VW.04,Ar.01,Wi.02,RC.05}).
There have also been relevant studies of entanglement dynamics in spin chains
(for instance \cite{Bos.04,AOF.04,SSL.04,HK.05,HgK.06,KRB.05}), which discuss
in particular the evolution of initial Bell states and the ensuing
``entanglement waves'' \cite{AOF.04}, non-ergodicity and dynamical phase
transitions starting with equilibrium states \cite{SSL.04}, decoherence waves
\cite{HK.05}, evolution in varying magnetic fields \cite{HgK.06}, generation of
cluster states \cite{KRB.05} as well as other issues.

In the present work we want to focus on a particular aspect, namely the
generation of entanglement in an interacting spin chain with fixed parameters
starting from an initially fully separable aligned state, and examine the
maximum entanglement that can be reached as a function of the anisotropy and
the uniform transverse magnetic field (control parameter). We will concentrate
here on the {\it global} entanglement between one qubit and the rest of the
chain and on the {\it pairwise} entanglement between neighboring qubits, within
the context of a cyclic $XY$ chain with nearest neighbor interactions
\cite{LSM.61}. Questions which immediately arise include the possible existence
of a threshold anisotropy for reaching maximum global entanglement
(saturation), the maximum pairwise entanglement that can be reached and, most
important, the behavior with the applied magnetic field. It will be shown that
contrary to what can be naively expected, the maximum global entanglement
reached is not a monotonous function of the aligning field, but exhibits
instead a typical {\it resonant behavior} for low anisotropies, with narrow
peaks located at characteristic field values, entailing a high sensitivity
suitable for entanglement control. The pairwise entanglement exhibits a more
complex resonant response, since it is affected by a competition between two
incompatible types (essentially of positive or negative spin parity). These
resonances {\it remain finite} in the limit of vanishing (but non-zero)
anisotropy in finite chains, considering sufficiently long time evolutions. On
the other hand, for large anisotropies they merge into a single broad maximum
centered at zero field, with global saturation reached within a field window.

Our results are based on a fully exact treatment of the finite $n$-spin chain
based on the Jordan-Wigner transformation \cite{LSM.61}, explicitly verified
for the case of two and three-qubit chains. The Hamiltonian and the
entanglement measures employed  are discussed in section II. Section III
contains the results, discussing first the two and three-qubit cases and then
the exact results for  general $n$-qubit chains. Finally, conclusions are drawn
in IV.

\section{Formalism}
We  consider $n$ qubits or spins in a cyclic chain interacting through an XY
nearest neighbor coupling, embedded in a uniform transverse magnetic field
\cite{LSM.61,S.99}. The Hamiltonian reads
\begin{subequations}\label{H1}
\begin{eqnarray}H&=&bS^z-\sum_{j=1}^n(v_xs^x_js^x_{j+1}+v_ys^y_js^y_{j+1})\\
&=&bS^z-\frac{1}{2}\sum_{j=1}^n (vs^+_js^-_{j+1}+gs^+_js^+_{j+1}+h.c.)\,,
\end{eqnarray}
\end{subequations}
where $S^z=\sum_{j=1}^n\!s^z_j$ is the total spin along the direction of the
magnetic field $b$, $v,g=(v_x\pm v_y)/2$ and $n+1\equiv 1$. We will consider
the evolution of the state which is initially fully aligned antiparallel to the
magnetic field,
\begin{equation}
|\Psi(t)\rangle=\exp[-iHt]|\!\!\downarrow\ldots\downarrow\rangle\,,\label{Psi}
\end{equation}
where $t$ denotes time over $\hbar$, and examine the emerging {\it global}
entanglement between one qubit and the rest of the chain, as well as the
{\it pairwise} entanglement between contiguous qubits, arising for non-zero
anisotropy $\gamma=g/v$ (for $g=0$ the initial state is an eigenstate of $H$
and hence no entanglement is generated).

Since we are dealing with a pure state, the first one is determined by the
entropy \cite{Be.96}
\begin{equation}
E_1=-{\rm Tr}\,\rho_1\log_2\rho_1\,,
\end{equation}
of the reduced {\it one-qubit} density $\rho_1={\rm Tr}_{n-1}\,\rho$, where
$\rho=|\Psi(t)\rangle\langle\Psi(t)|$ is the full density matrix, with $E_1=0$
for $\rho_1$ pure ($\rho_1^2=\rho_1$) and $E_1=1$ (maximum) for $\rho_1$ fully
mixed.  The second one is the entanglement of formation \cite{Be.96} of the
adjacent {\it pair} density $\rho_{2}={\rm Tr}_{n-2}\,\rho$, which can be
calculated as \cite{W.98}
\begin{equation}
E_2=-\!\sum_{\nu=\pm} q_\nu\log_2 q_\nu\,,
\end{equation}
where $q_{\pm}=(1\pm\sqrt{1-C_2^2})/2$ and
\begin{equation}
C_2={\rm Max}[2\lambda_{m}-{\rm Tr}\,R,0],\;\;
 R=\sqrt{\rho_2\tilde{\rho}_2}\,,\label{C2}
\end{equation}
is the {\it concurrence} \cite{W.98}, with $\lambda_m$ the greatest eigenvalue
of $R$ and $\tilde{\rho}_2=4s^y_js^y_{j+1}\rho_2^*s^y_{j+1}s^y_j$ the
spin-flipped density. It satisfies $0\leq C_2\leq 1$. Since tracing out qubits
of a subsystem can be considered a LOCC (local operations and classical
communication) transformation, it cannot increase entanglement \cite{Be.96} and
hence $E_{2}\leq E_1$, with $E_2=E_1$ for a pure two qubit state (in which case
$q_\pm$ become the eigenvalues of $\rho_1$).

As $E_2$ is just an increasing function of $C_2$,  pairwise entanglement is
usually directly measured through the latter, which is more suitable for
analytic description. The corresponding measure of the global
$E_1$ entanglement is the square root of the {\it one-tangle} \cite{W.98},
 \begin{equation}
C_1=2\sqrt{{\rm Det}\,\rho_1}=\sqrt{2(1-{\rm Tr}\,\rho^2)}\,,\label{C1}
 \end{equation}
which coincides with $C_2$ for a pure two qubit state and satisfies $C_1\geq
C_2$ in the general case (actually the more general inequality $C_i\geq
\sqrt{\sum_{j\neq i}C_{ij}^2}$, with $C_{ij}$ the concurrence of the $(i,j)$
pair and $C^2_i$ the one-tangle of qubit $i$, conjectured in \cite{CKW.00}, was
recently proven \cite{OV.06}). Both $E_1$ and $C_1$ are measures of the
disorder associated with $\rho_1$ and are hence increasing functions of one
another.

Due to the symmetries of $H$ and the present initial state, $|\Psi(t)\rangle$
will be invariant under translation $(j\rightarrow j+1)$ and inversion
($j\rightarrow n+1-j$), and will have {\it positive spin parity}
$P=\exp[i\pi(S^z+n/2)]$, as this quantity is preserved by $H$ ($[H,P]=0$). The
reduced density $\rho_{S}={\rm Tr}_{n-S}\rho$ of {\it any} subsystem $S$ will
then depend just on the distance between its components and will {\it commute}
with the subsystem parity $P_S=\prod_{j\in S}\exp[i\pi(s_z^j+1/2)]$, as the
reduction involves just diagonal elements in the rest of the chain. In the case
of $\rho_1$, this implies that it will be the same for all qubits and {\it
diagonal} in the standard basis $|\!\!\uparrow\rangle,|\!\!\downarrow\rangle$
of $s^z$ eigenstates:
\begin{equation}
\rho_1=\left(\begin{array}{cc}p(t)&0\\0&1-p(t)\end{array}\right)\,,
\end{equation}
where $p(t)$ represents the one-qubit spin flip probability
\begin{equation}
p(t)=\langle s^z_j\rangle_t+1/2=\langle S^z\rangle_t/n+1/2\,,
\end{equation}
(here $\langle O\rangle_t\equiv \langle\Psi(t)|O|\Psi(t)\rangle$ and spin
operators are considered dimensionless). Hence,
\begin{equation}
C_1(t)=2\sqrt{p(t)[1-p(t)]}\,,\label{C1t}
\end{equation}
with $C_1(t)=1$ when $p(t)=1/2$.

The same symmetries lead to a pair density of the form
\begin{equation}
\rho_{2}=\left(\begin{array}{cccc}p_{1}(t)&0&0&\alpha^*(t)
\\0&p_{2}(t)&\beta(t)&0\\0&\beta(t)&p_2(t)&0\\
\alpha(t)&0&0&p_3(t)\end{array}\right)\,,
\end{equation}
in the standard basis $|\!\!\uparrow\uparrow\rangle,
|\!\!\uparrow\downarrow\rangle,|\!\!\downarrow\uparrow\rangle,
|\!\!\downarrow\downarrow\rangle$, where
$p_{1}(t) +2p_{2}(t)+p_{3}(t)=1$, $p_{1}(t)+p_{2}(t)=p(t)$ and
\begin{subequations}
\label{alp}
\begin{eqnarray}
\alpha(t)&=&\langle s^+_js^+_{j+1}\rangle_t\,,
\;\;\beta(t)=\langle s^+_js^-_{j+1}\rangle_t\,,\\
p_1(t)&=&\langle (s^z_j+1/2)(s^z_{j+1}+1/2)\rangle_t\,,
\end{eqnarray}
\end{subequations}
for adjacent qubits.  Eq.\ (\ref{C2}) becomes then
\begin{equation}
C_{2}(t)=2\,{\rm Max}\,
[|\alpha(t)|-p_{2}(t),|\beta(t)|-\sqrt{p_{1}(t)p_{3}(t)},0]\,, \label{C2t}
\end{equation}
where only one of the entries can be positive (this follows from the positivity
of $\rho_2$, which requires $|\alpha(t)|\leq \sqrt{p_1(t)p_3(t)}$,
$|\beta(t)|\leq p_2(t)$). Two kinds of pairwise entanglement can therefore
arise: type I ($|\alpha(t)|>p_{2}(t)$) and  type II
($|\beta(t)|>\sqrt{p_{1}(t)p_{3}(t)}$), which cannot coexist and can then be
present just at {\it different} times, and which stem from the positive (I) and
negative (II) parity sectors of $\rho_2$.

The eigenvalues of $H$ and the entanglement of its eigenstates are obviously
independent of the sign of $g$, and for even chains also of the sign of $v$, as
for $n$ even it can be changed by a transformation $s_j^{x,y}\rightarrow
(-1)^js_j^{x,y}$. Due to time reversal symmetry, the emerging entanglement in
even chains will then be also independent of the sign of $b$, while in odd
chains that for $(-b,v)$ will coincide with that for $(b,-v)$. We will then set
in what follows $v\geq 0$, $g\geq 0$, and consider both signs of $b$.

\section{Results}
\subsection{Two qubit case}
\begin{figure}[t]

 \centerline{\scalebox{0.7}{\includegraphics{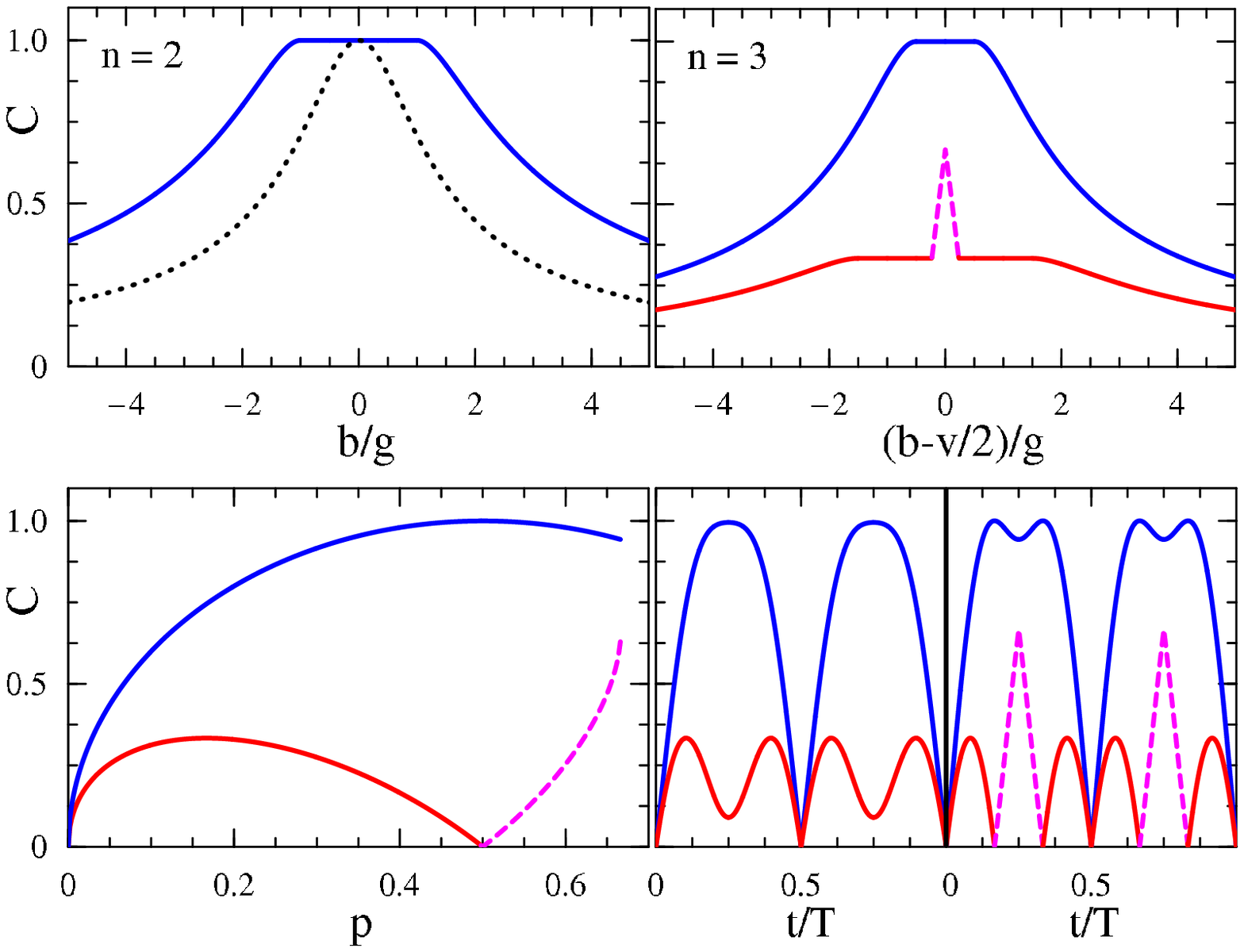}}}
\caption{(Color online). Top: Left: Maximum entanglement (measured by the
concurrence) reached by the two qubit chain as a function of the (scaled)
magnetic field for an initially aligned state. The dotted line depicts the
concurrence of the Hamiltonian eigenstates. Right: Maximum global concurrence
$C_1^m$ between one-qubit and the rest (upper curve, in blue) and maximum
pairwise concurrence $C_2^m$ (lower curve, in red+dashed pink) in the three
qubit system, in terms of the (shifted+scaled) magnetic field. $C_2^m$ exhibits
a sharp type II resonance at $b=v/2$. Bottom: Left: Plot of $C_1$ and $C_2$ in
the three qubit chain in terms of the spin flip probability $p$ ($0\leq p\leq
2/3$). Right: The temporal evolution of $C_1$ and $C_2$ in the three qubit
chain at the $C_2^m$ plateau ($b=v/2\pm 0.6 g$, left) and at resonance
($b=v/2$, right). $T=2\pi/\lambda$ is the period. Type I (II) sectors in $C_2$
are depicted in solid red (dashed pink) lines.}
 \label{f1}\vspace*{0.cm}
\end{figure}

Let us first analyze this simple situation, which nonetheless provides already
some insight on the behavior for general $n$. Here $C_1=C_2$ $\forall\, t$. The
evolution subspace is spanned by the states $|\!\!\downarrow\downarrow\rangle$,
$|\!\!\uparrow\uparrow\rangle$, and the pertinent eigenstates of $H$ are
$|\pm\rangle=u_{\mp}|\!\!\downarrow\downarrow\rangle\mp u_{\pm}
|\!\!\uparrow\uparrow\rangle$, with energies $E_{\pm}=\pm \lambda$, where
$u_{\pm}=\sqrt{(\lambda\pm b)/(2\lambda)}$ and $\lambda=\sqrt{b^2+g^2}$. The
state (\ref{Psi}) will then be independent of $v$ and given by
\begin{eqnarray}
|\Psi(t)\rangle&=&\sum_{\nu=\pm}e^{-iE_\nu t}
\langle\nu|\!\!\downarrow\downarrow\rangle|\nu\rangle \nonumber\\
&=&(\cos\lambda t+i{\textstyle\frac{b}{\lambda}}\sin\lambda t)
|\!\!\downarrow\downarrow\rangle
+i{\textstyle\frac{g}{\lambda}}\sin\lambda t
|\!\!\uparrow\uparrow\rangle\,,\label{Psi2}
\end{eqnarray}
so that the spin-flip probability $p(t)$ is
\begin{equation}
p(t)={\frac{g^2}{b^2+g^2}}\sin^2\lambda t\,.\label{p2t}
\end{equation}
Its maximum $p_m=g^2/(b^2+g^2)$ is thus {\it a Lorentzian of width} $g$
centered at $b=0$, satisfying $p_m\geq 1/2$ if $|b|\leq g$. Hence, for {\it
any} $g>0$ the system will always reach {\it maximum entanglement} $C_1=1$
within the field window $|b|\leq g$, at times $t_m$ such that $p(t_m)=1/2$,
where Eq.\ (\ref{Psi2}) becomes a type I Bell state:
\[|\Psi(t_m)\rangle=\pm i(|\!\!\uparrow\uparrow\rangle+
e^{\pm i\phi}|\!\!\downarrow\downarrow\rangle)/\sqrt{2}\,,\;\;\cos\phi=b/g\,.\]
The maximum concurrence reached (Fig.\ \ref{f1}) is then
\begin{equation}
C_1^m=C_2^m=\left\{\begin{array}{lr}1\,,&|s|\leq 1\\
\frac{2|s|}{s^2+1}\,,&|s|\geq 1\end{array}\right.\,,\;\;s=b/g\,,
\label{C12}
\end{equation}
which is {\it higher} than the concurrence $C^{\pm}=g/\lambda$ of the
Hamiltonian eigenstates $\forall$ $b\neq 0$, becoming $\approx 2g/|b|$ for
$|b|\gg g$. $C_1(t)$ will follow the evolution of $p(t)$ if $p_m\leq 1/2$
($|s|\leq 1$), but will develop saturated maxima plus an intermediate minima
when $p_m>1/2$.

We also note that for $b=0$, i.e., where the gap $E_+-E_-=2\lambda$ is minimum
and vanishes for $g\rightarrow 0$, maximum entanglement can in principle be
attained for {\it any} $g>0$. In this case the eigenstates $|\pm\rangle$ become
{\it independent} of $g$ and maximally entangled, and none of them approaches
the aligned initial state for $g\rightarrow 0$ (in contrast with the behavior
for $b\neq 0$). The initial state becomes then equally distributed over both
eigenstates ($u_{\pm}=1/\sqrt{2}$) $\forall$ $g>0$, implying
$|\Psi(t)\rangle=\cos gt|\!\!\downarrow\downarrow\rangle+i\sin gt
|\!\!\uparrow\uparrow\rangle$. Hence, in this case the only limit for reaching
maximum entanglement ($\sin^2 gt=1/2$) for arbitrarily small but non-zero $g$
is the long waiting time ($t_m=\pi/(4g)$). We will see that an analogous
situation will occur for any $n$ at particular field values.

\subsection{Three qubit case}
For $n=3$, the evolution subspace is still two-dimensional and spanned by
$|\!\!\downarrow\downarrow\downarrow\rangle$ and the $W$-state \cite{DC.00}
$|W\rangle\equiv(|\!\!\downarrow\uparrow\uparrow\rangle+
|\!\!\uparrow\downarrow\uparrow\rangle+
|\!\!\uparrow\uparrow\downarrow\rangle)/\sqrt{3}$, which for $g=0$ have
energies $-3b/2$ and $b/2-v$. The coupling induced by $g$ leads to eigenstates
$|\pm\rangle=u_{\mp}|\!\!\downarrow\downarrow\downarrow\rangle\mp
u_{\pm}|W\rangle$ with energies $E_{\pm}=\varepsilon\pm\lambda$, where
$u_{\pm}=\sqrt{[\lambda\pm (b-v/2)]/(2\lambda)}$, $\varepsilon=-(b+v)/2$  and
$\lambda=\sqrt{(b-v/2)^2+3g^2/4}$. We then obtain
\[|\Psi(t)\rangle=e^{-i\varepsilon t}[(\cos\lambda t+i
{\textstyle\frac{b-v/2}{\lambda}}\sin\lambda t)
|\!\!\downarrow\downarrow\downarrow\rangle
+i{\textstyle\frac{\sqrt{3}g}{2\lambda}}\sin\lambda t|W\rangle]\]
 which leads to
 \begin{equation}
p(t)=\frac{g^2}{2[(b-v/2)^2+3g^2/4]}\sin^2\lambda t\label{p3t}\,.
\end{equation}
Its maximum $p_m=g^2/(2\lambda^2)$ is again a Lorentzian of width proportional
to $g$ but centered at $b=v/2$ due to the hopping term, where $p_m=2/3$ (the
value at the $W$-state), with $p_m\geq 1/2$ for $|b-v/2|\geq g/2$. Hence, for
{\it any} $g\neq 0$ there is again a field interval where {\it maximum} $E_1$
entanglement is attained. The maximum of $C_1(t)$ (Fig.\ \ref{f1}, top right)
is then
\begin{equation}
C_1^m=\left\{\begin{array}{lr}1\,,&|s|\leq 1/2\\
\frac{\sqrt{2s^2+1/2}}{s^2+3/4}\,,&|s|\geq 1/2\end{array}\right.\,,\;
s=(b-v/2)/g\,.\label{C13}
\end{equation}
For $|b|\gg v,g$, $C_1^m\approx \sqrt{2}g/|b|$, an asymptotic result which
turns out to be {\it valid $\forall$ $n\geq 3$}. The evolution of $C_1(t)$
remains qualitatively similar to that for $n=2$. Note also that for $b=v/2$,
i.e., where the gap $2\lambda$ is minimum and vanishes for $g=0$, maximum $C_1$
is again reached for any $g>0$, the situation being similar to that for $n=2$
at $b=0$.

The behavior of the pairwise entanglement is, however, more complex. The
$W$-state contains type II pairwise entanglement, but $|\Psi(t)\rangle$ will
first develop that of type I, so that transitions between both types
can be expected to occur in the evolution for large $g$. From the expression of
$|\Psi(t)\rangle$ we obtain $|\alpha(t)|=\sqrt{p(t)(2-3p(t))}/2$,
$p_{2}(t)=p_{1}(t)=\beta(t)=p(t)/2$, so that Eq.\ (\ref{C2t}) becomes
\begin{equation}
C_{2}(t)=|\sqrt{p(t)[2-3p(t)]}-p(t)|\,,\label{C23}
\end{equation}
which corresponds to type I (II) for $p(t)<1/2$ ($>1/2$). It is thus a
non-monotonous function of $p\equiv p(t)$ (left bottom panel in Fig.\
\ref{f1}), having a maximum at $p=1/6$ (where $C_2=1/3$), vanishing at the
``critical'' value $p=1/2$ (where $C_1$ is maximum) and increasing again for
$p>1/2$ up to its absolute maximum at the endpoint $p=2/3$, where $C_2=2/3$
(i.e., the value at the $W$-state). Hence, saturation ($C_2=1$) cannot be
reached. Moreover, it is verified that $C_{2}(t)/C_1(t)\leq 1/\sqrt{2}$ (the
maximum ratio allowed by the generalized inequality \cite{CKW.00} for
$C_{12}=C_{13}$), the maximum reached for $p\rightarrow 0$ or $p\rightarrow
2/3$. The evolution of $C_2(t)$ will then not follow that of $p(t)$ or $C_1(t)$
if $p_m>1/6$, developing for $p_m<1/2$ a minimum when $p(t)$ is maximum, which
will evolve into two vanishing points plus a type II maximum if $p_m>1/2$ (see
right bottom panel in Fig.\ \ref{f1}). The maximum of $C_{2}(t)$ is then
\begin{equation}
C_{2}^m=\left\{\begin{array}{lc}
\frac{1/2-|s|}{s^2+3/4}\,, &|s|\leq s_c\\
 1/3\,,&s_c\leq |s|\leq 3/2\\
\frac{|s|-1/2}{s^2+3/4}\,, &|s|\geq 3/2\\
\end{array}\right.
\,,\;s=\frac{b-v/2}{g} \label{C23m}
\end{equation}
where $s_c=\sqrt{3}-3/2\approx 0.23$ determines the second point where
$C_2=1/3$ and encloses the region of dominant type II entanglement. It then
exhibits {\it a sharp type II peak at $b=v/2$}, above a strict {\it type I
plateau} (see Fig.\ \ref{f2}). Note that at $b=v/2$, $C_2^m=2/3$ for {\it any}
$g>0$, as in this case the system will always reach the $W$-state if the
waiting time is sufficiently long ($t_m=\pi/(\sqrt{3}g)$). For $|b|\gg v,g$,
$C_{2}^m\approx g/|b|\approx C_1^m/\sqrt{2}$, an asymptotic result {\it which
is again  valid $\forall$ $n\geq 3$}.

\subsection{General $n$}
By means of the Jordan-Wigner transformation \cite{LSM.61}, we may exactly
convert the Hamiltonian (\ref{H1}) within a fixed spin parity subspace ($P=\pm
1$) to a quadratic form in fermion operators $c_j^\dagger,c_j$, defined by
$c^\dagger_{j}=s^+_j\exp[-i\pi\sum_{l=1}^{j-1}s^+_{l}s^{-}_{l}]$.
 For a finite cyclic chain with positive parity
$P=1$, the result for $H'=H+bn/2$ is
\begin{subequations}
\begin{eqnarray}
H'&=&\sum_{j=1}^n bc^\dagger_jc_j-({\textstyle\frac{1}{2}}-\delta_{jn})
(vc^\dagger_jc_{j+1}+gc^\dagger_{j}c^\dagger_{j+1}+h.c.)\label{Hf1}\\
&=&\sum_{k}(b-v\cos\omega_k)c'^\dagger_kc'_k-{\textstyle\frac{1}{2}}
g\sin\omega_k(c'^\dagger_kc'^\dagger_{-k}+c'_{-k}c'_k)\,,
 \label{Hf2}\end{eqnarray}
 \end{subequations}
where the fermion operators $c'_k,c'^\dagger_k$ are related to
$c_j,c^\dagger_j$ by a finite Fourier transform
\[c^\dagger_j={\textstyle\frac{e^{i\pi/4}}{\sqrt{n}}}
 \sum_{k}e^{i\omega_k j}c'^\dagger_k,\;\;\omega_k=2\pi k/n\,,\]
with $k$ {\it half-integer} for the present cyclic conditions:
$k=-\frac{n-1}{2},\ldots,\frac{n-1}{2}$ for $n$ even and
$k=-\frac{n}{2}+1,\ldots,\frac{n}{2}$ for $n$ odd. We then obtain the diagonal
form
 \begin{eqnarray}H'&=&
\sum_{k}\lambda_k a^\dagger_k a_k-{\textstyle\frac{1}{2}}
[\lambda_k-(b-v\cos\omega_k)]\,,\nonumber\\
\lambda_k&=&\sqrt{(b-v\cos\omega_k)^2+g^2\sin^2\omega_k}\,,\label{lj}
 \end{eqnarray}
by a means of a BCS-like transformation $c'^\dagger_k=
u_ka^\dagger_k+v_ka_{-k}$, $c'_{-k}=u_ka_{-k}-v_ka^\dagger_k$ to quasiparticle
fermion operators $a^\dagger_k,a_k$, with $u_k^2,v_k^2=[\lambda_k\pm
(b-v\cos\omega_k)]/(2\lambda_k)$. The quasiparticle energies (\ref{lj}) are
two-fold degenerate ($\lambda_k=\lambda_{-k}$) except for $k=n/2$ for $n$ odd.

We can now determine the exact evolution for any $n$. In the Heisenberg
representation ($dO/dt=i[H,O]$), we have
$a^\dagger_k(t)=e^{i\lambda_kt}a^\dagger_k(0)$,
 $a_k(t)=e^{-i\lambda_k t}a_k(0)$, and the ensuing contractions
\[\langle a^\dagger_k(t)a_k(t)\rangle_0=v_k^2\,,\;\;\;
\langle a^\dagger_k(t)a^\dagger_{-k}(t)\rangle_0=-u_kv_k e^{2i\lambda_k t}\,,\]
with respect to  the present initial state (vacuum of the operators $c,c'$).
The average of any operator can now be evaluated by substitution and use of
Wick's theorem \cite{RS.80}.

\subsubsection{Evaluation of $p(t)$ and $C_1(t)$}
The one-qubit spin flip probability becomes
\begin{equation}
p(t)=\langle c^\dagger_j(t)c_j(t)\rangle_0
=\frac{2}{n}{\sum_k}'\frac{g^2\sin^2\omega_k}{\lambda_k^2}\sin^2\lambda_kt
 \label{pnt}\,,\end{equation}
where $\sum'_k\equiv\sum_{k=1/2}^{[n/2]-1/2}$ ($[n/2]$ denotes integer part).
For $n=2,3$ the sum in (\ref{pnt}) reduces to a single term ($k=1/2$, with
$\omega_k=\pi/2$ and $\pi/3$ respectively) and we recover {\it exactly} Eqs.\
(\ref{p2t}) and (\ref{p3t}).

For $n\geq 4$, the evolution of $p(t)$ will be in general quasiperiodic. Its
upper envelope can nevertheless be obtained setting $\sin^2\lambda_k t=1$
$\forall$ $k$ in (\ref{pnt}):
\begin{equation}
p(t)\leq p_m=\frac{2}{n} {\sum_k}'
\frac{g^2\sin^2\omega_k}{(b-v\cos\omega_k)^2+g^2\sin^2\omega_k}
\label{pmast}\,,
\end{equation}
the maximum of $p(t)$ lying arbitrarily close to $p_m$ for sufficiently long
time intervals (except for rational ratios $\lambda_k/\lambda_{k'}$). For low
$g\ll v$,  $p_m$  {\it will then exhibit $[n/2]$ peaks, located at}
\begin{equation}
b=b_k\equiv v\cos\omega_k,\;\;\;k={\textstyle\frac{1}{2},\ldots,
[\frac{n}{2}]-\frac{1}{2}}\,,
\end{equation}
(i.e. $\omega_k=\pi/n,3\pi/n,\ldots,(2[n/2]-1)\pi/n)$, which are the fields
where the quasiparticle energies $\lambda_{\pm k}$ are minimum and vanish for
$g\rightarrow 0$. Hence, they are located symmetrically around $b=0$ for even
$n$ ($b_{[n/2]-k}=-b_k$), with a peak at $b=0$ ($k=n/4$) for $n/2$ odd, but
asymmetrically for odd $n$. Moreover, while for $b\neq b_k$, $p_m\propto g^2$,
vanishing for $g\rightarrow 0$, {\it at $b=b_k$ $p_m$ remains finite $\forall$
$g\neq 0$, with $p_m\rightarrow 2/n$ for $g\rightarrow 0$} (Eq.\ \ref{pmast}).
This implies
\begin{equation}
C_1^m\rightarrow {\textstyle 2\sqrt{\frac{2}{n}(1-\frac{2}{n})}}\,,
\label{C1m}
\end{equation}
at $b=b_k$ for $g\rightarrow 0$ and $n\geq 4$ (and $C_1^m\rightarrow 1$ for
$n=2,3,4$ as in these cases $2/n\geq 1/2$). Thus, by adjusting the field it is
always possible to achieve, in principle, {\it finite} $E_1$ entanglement even
for arbitrarily low (but non-zero) values of $g$. The effect of low
anisotropies is just to determine the {\it width} of these peaks, given by
$\approx g|\sin\omega_k|$ in $p_m$, which increases as $g$ increases or as
$|b_k|$ decreases.

The evolution at $b=b_k$ becomes purely harmonic for $g\rightarrow 0$, with
\begin{equation}
p(t)\rightarrow{\textstyle\frac{2}{n}}\sin^2\lambda_kt\,,\;\lambda_k=
g\sin\omega_k\,.\label{ptg}
\end{equation}
The maximum of $p(t)$ is first reached at $t_k=\pi/(2g\sin\omega_k)$, so that
the smaller the value of $g$ (or $\omega_k$), the longer it will take to reach
the maximum. In this sense, while the maximum entanglement reached in an
unbounded time interval is not a continuous function of $g$ for $g\rightarrow
0$ at $b=b_k$, that reached  in a {\it finite} interval $[0,t_f]$ will actually
vanish for $g\rightarrow 0$ also at $b=b_k$, in agreement with the result for
$g=0$, becoming lower than (\ref{C1m}) if $t_f<t_k$.

The situation at the resonances $b=b_k$ is thus similar to that encountered for
$n=2$ at $b=0$ or for $n=3$ at $b=v/2$. At $b=b_k$ the energy gap $2\lambda_k$
between positive parity states with the pair $(k,-k)$ occupied and empty (in
particular that between the quasiparticle vacuum $|0_q\rangle$ and the state
$a^\dagger_ka^\dagger_{-k}|0_q\rangle$) is minimum, {\it vanishing} for
$g\rightarrow 0$ (level crossings). Due to these degeneracies, at $b=b_k$ the
aligned state is not approached by any of the Hamiltonian eigenstates for
$g\rightarrow 0$, remaining distributed over essentially two eigenstates. The
previous limits (\ref{C1m})-(\ref{ptg}) can then be directly derived from Eq.\
(\ref{Hf2}), where for $g\rightarrow 0$ and $b=b_k$, we may conserve just the
$\pm k$ terms in the $g$-interaction. The evolution subspace in this limit is
then spanned by the original fermionic vacuum $|0\rangle$ (the present initial
state) and the two particle state
$|k,-k\rangle=c'^\dagger_kc'^\dagger_{-k}|0\rangle$, with $g$-independent
eigenstates $|\pm\rangle=(|0\rangle\mp|k,-k\rangle)/\sqrt{2}$ of perturbed
energies $\pm g\sin\omega_k$ (i.e., $\pm \lambda_k$). We then obtain (omitting
a global phase)
\begin{equation}
|\Psi(t)\rangle\rightarrow\cos \lambda_k t|0\rangle+i\sin\lambda_k
t|k,-k\rangle\,, \label{apro}
\end{equation}
for the fermionic $|\Psi(t)\rangle$, which leads immediately to Eq.\
(\ref{ptg}). The factor $2/n$ is just the average occupation $\langle
c^\dagger_j c_j\rangle=\sum_{k'}\langle c'^\dagger_{k'}c'_{k'}\rangle/n$ in the
state $|k,-k\rangle$.

As $g$ increases, the resolutions of the individual peaks diminish, merging
eventually into a single broad peak centered at $b\approx 0$. Since the
separation between maxima is $\delta b\approx (2\pi v/n)|\sin\omega_k|$, we
have the approximate bound $g\alt\pi v/n$ for  visible individual peaks. On the
other hand, it is to be noticed that for $n\geq 5$ maximum $E_1$ entanglement
can be reached only above a certain {\it threshold} value $g_c$ of $g$ (and
then within a certain field window), with $g_c\leq v$ $\forall$ $n$ since at
$b=0$ and $g=v$ we have exactly $p_m=(2/n)\sum'_k\sin^2\omega_k=1/2$ for {\it
any} $n$. In fact, $g_c\approx v$ for large $n$. For $g\gg (v,b)$,
$p_m\rightarrow 1$ $(1-1/n)$ for $n$ even (odd), so that saturation in $C_1$ is
always reached. Finally, for large fields $|b|\gg v,g$,
\begin{equation}
p_m\approx \frac{2g^2}{n b^2}{\sum_k}'\sin^2\omega_k
=\frac{g^2}{2b^2},\;\;n\geq 3\,,
\end{equation}
implying $C_1^m\approx \sqrt{2}g/|b|$. This asymptotic result is {\it
independent} of $n$ (for $n\geq 3$) and coincident with the result previously
obtained for $n=3$.

\begin{figure}[t]

\centerline{\scalebox{0.7}{\includegraphics{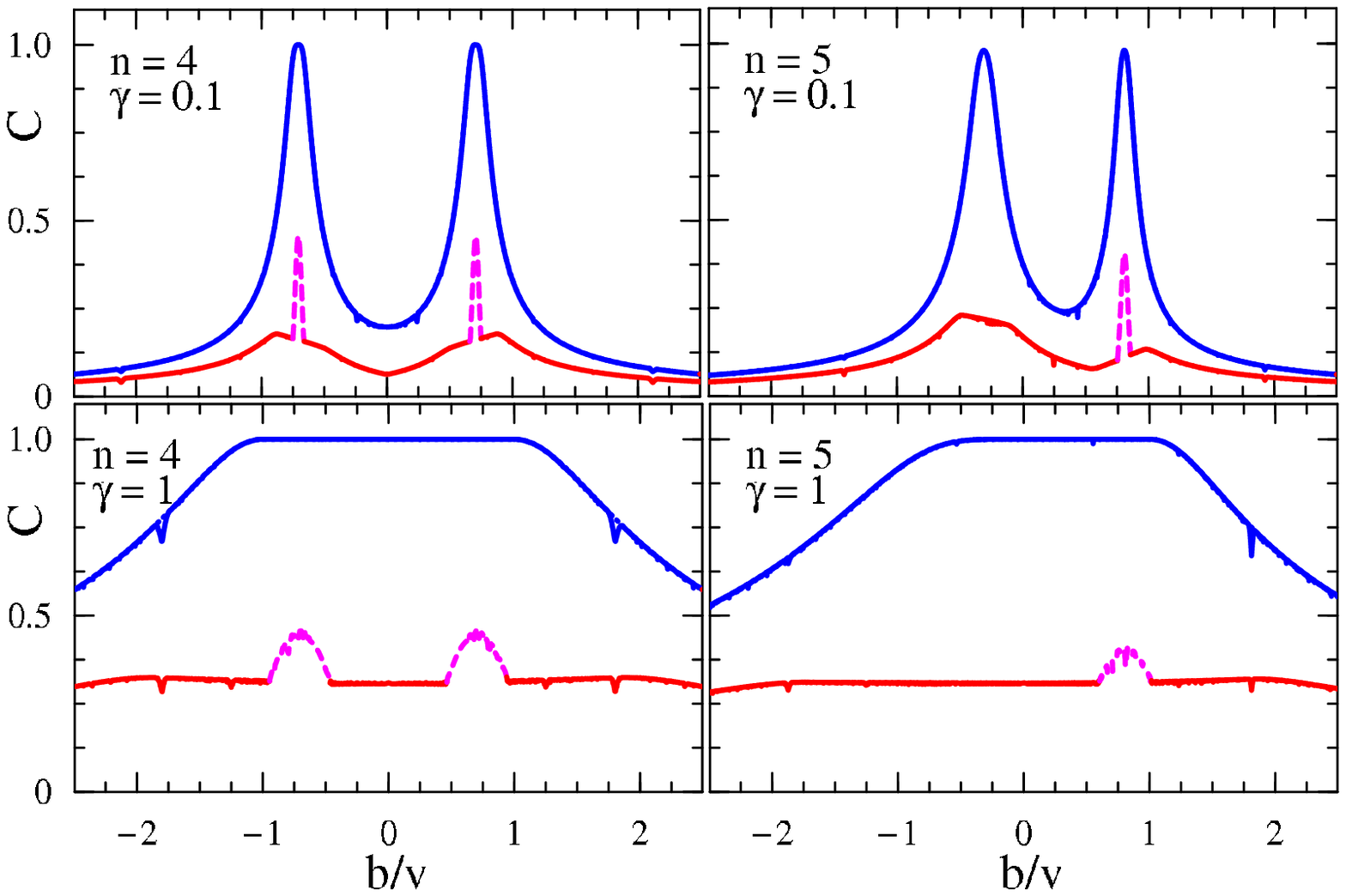}}}
\caption{(Color online). Maximum concurrence between one qubit and the rest
(upper blue curves) and between adjacent qubits (lower red+dashed pink curves),
reached in the four (left) and five (right) qubit chains for two different
anisotropies $\gamma=g/v$ (type II sectors in $C_2$ depicted again with dashed
pink lines). For $n=4$ the peaks in the global concurrence at $b/v=\pm
1/\sqrt{2}$ are no longer resolved for $\gamma\geq 1$, but remain in the
pairwise concurrence. For $n=5$, the resonances are located at
$b/v=(1\pm\sqrt{5})/4$ and merge again in a saturated maximum for $\gamma\geq
1$, while the pairwise concurrence presents a type II resonance just at the
second peak, which again remains visible for large $\gamma$. Dotted lines in
the upper curves depict results obtained with the upper envelope (\ref{pmast}),
and are almost coincident with the numerically obtained maximum in the interval
$0\leq vt\leq 40$. See text for more details.}
 \label{f2}\vspace*{0.cm}
\end{figure}

\begin{figure}[t]

 \centerline{\scalebox{0.7}{\includegraphics{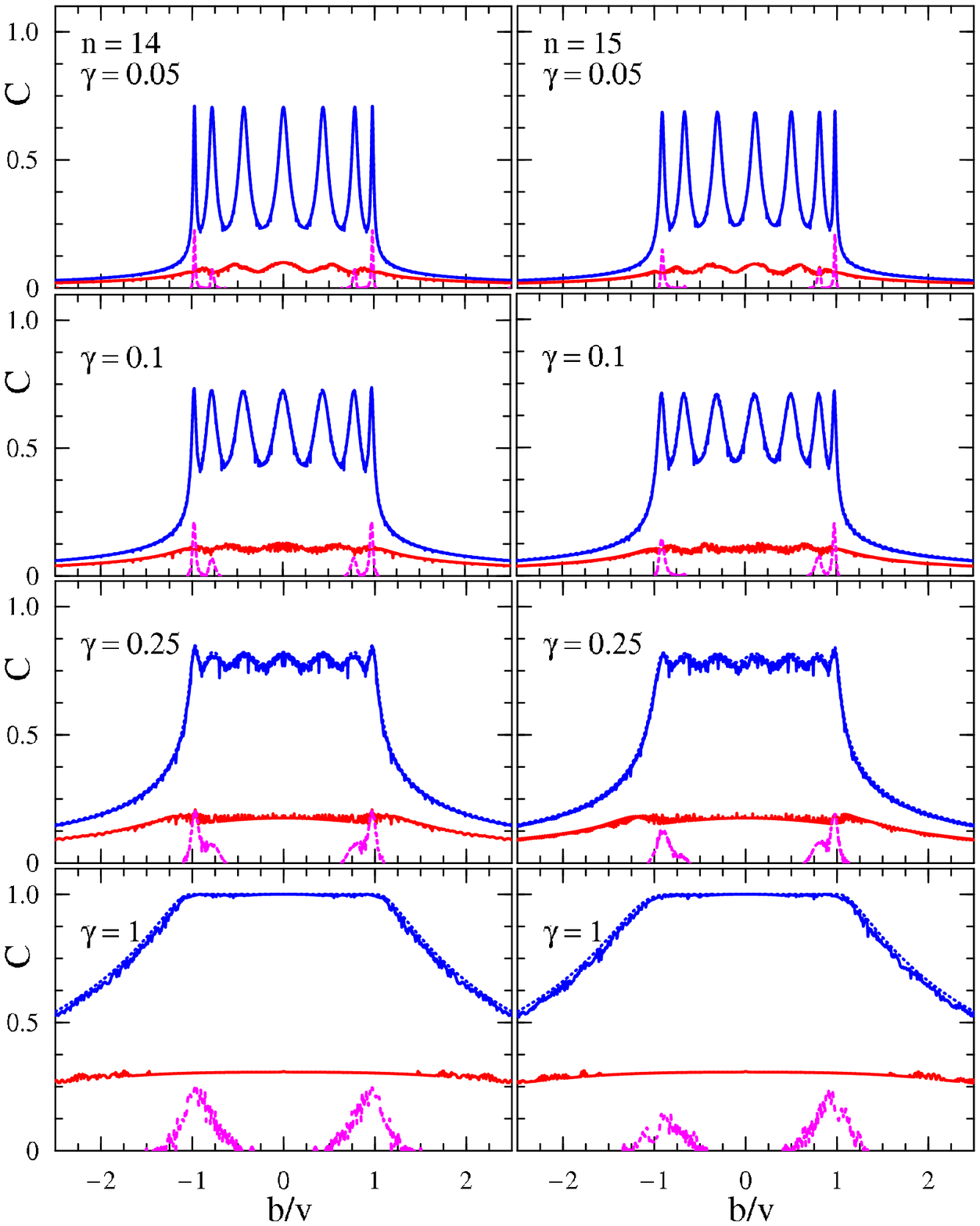}}}
\caption{(Color online) Maximum concurrence between one qubit and the rest of
the chain (upper blue lines) and between adjacent qubits (lower red+dashed pink
lines) in a $n=14$ (left) and $n=15$ (right) qubit chain for different
anisotropies, reached in an interval $0\leq vt\leq 180$. The dashed pink lines
depict the maximum of the type II pairwise concurrence, which becomes now lower
than the type I plateau for $\gamma\agt 0.25$. Results for $C_1$ obtained with
the upper bound (\ref{pmast}) are also depicted (dotted lines, almost
overlapping with the blue solid lines). The peaks in $C_1$ are visible for
$\gamma\alt 0.4$, and saturation ($C_1=1$) is reached for $\gamma\agt 0.92$.}
 \label{f3}\vspace*{0.cm}
\end{figure}

Results for $n=4,5$ and $14,15$ are shown in Figs.\ \ref{f2} and \ref{f3}. For
$n=4$, the resonances are located at $b_k=\pm v/\sqrt{2}$, with $p_m\geq 1/2$
(and hence $C_1^m=1$) for $|b^2-v^2/2|\leq g^2/2$. This determines two
saturated plateaus in $C_1^m$ centered at $b=b_k$ for $g<v$, which merge into a
{\it single} plateau centered at $b=0$ for $g>v$. For $n=5$ the peaks are
located at $b_k=v(1\pm\sqrt{5})/4\approx 0.81,-0.31$, where $C_1^m\rightarrow
2\sqrt{6}/5\approx 0.98$ for $g\rightarrow 0$ (Eq.\ \ref{C1m}). Saturation is
reached only for $g/v\agt 0.67$, initially just at the right peak, although for
$g>v$, $C_1^m$ exhibits again a saturated plateau covering $b=0$. For $n=14$
(15), $C_1^m\rightarrow 0.7$ (0.68) at the seven peaks for $g\rightarrow 0$,
and saturation is reached for $g\agt 0.92$.

The small or tiny dips in the numerical result for $C_1^m$ that can be seen in
Figs.\ 2 and 3 arise due to the occurrence of rational ratios between the
quasiparticle energies $\lambda_k$ at particular values of $b/v$, in which case
the maximum of $p(t)$ can be lower than the smooth upper envelope
(\ref{pmast}). For instance, for $n=4$ the ratio of the two distinct energies
$\lambda_{1/2}$, $\lambda_{3/2}$ becomes $2$ at $|b|/v=\sqrt{2}(5\pm
\sqrt{16-9\gamma^2})/6$ (provided $\gamma<4/3$), where the maximum reached by
$p(t)$ is just $(4/5)p_m$ (20\% reduction). A reduction in the maximum of
$p(t)$ will also occur in the vicinity of these values of $|b|/v$ for finite
time intervals. This effect gives rise to the noticeable dip in $C_1^m$ at
$|b|/v\approx 1.8$ for $\gamma=1$ (the other value $|b|/v\approx 0.55$ lies
within the plateau region and its effect on $C_1^m$ is unobservable) and to
those at $|b|/v\approx 0.24$ and $\approx 2.12$ for $\gamma=0.1$.

It should be also mentioned that for short times $\lambda_k t\ll 1$ $\forall
k$, $p(t)$ becomes independent of $n$, its series expansion of order $m$
remaining stable for $n>m$. For instance, up to $O((\lambda_kt)^4)$ in $p(t)$,
we obtain, for $n\geq 5$,
 \begin{eqnarray}
p(t)&\approx& {\textstyle\frac{1}{2}g^2t^2[1-\frac{1}{12}t^2(v^2+4b^2+3g^2)]}
\,,\nonumber\\
C_1(t)&\approx&\sqrt{2}gt[1-{\textstyle\frac{1}{24}}t^2(v^2+4b^2+9g^2)/24]\,.
 \nonumber\end{eqnarray}
It is thus seen that for $g\gg (b,g)$ and $n\agt 8$, $p(t)$ exhibits an initial
peak at $t\approx 1.92/g$, where $p(t)\approx 0.7$, with $p(t)\geq 1/2$ for
$1.2\alt g t\alt 2.75$, so that in this limit saturation in $C_1$ is rapidly
reached (see Fig.\ \ref{f5}). The initial peak in $C_1$ can be correctly
predicted by its $7^{\rm th}$ order expansion.

\subsubsection{Evaluation of $C_2(t)$}
Let us now examine the pairwise concurrence. The relevant elements
(\ref{alp}) of the adjacent pair density are
\begin{eqnarray}
\beta(t)&=&\langle c^\dagger_j(t)c_{j+1}(t)\rangle_0=
\frac{2}{n}{\sum_{k}}'\frac{g^2\cos\omega_k\sin^2\omega_k}{\lambda_k^2}
\sin^2\lambda_kt\,,\nonumber\\
\alpha(t)&=&\langle c^\dagger_j(t)c^\dagger_{j+1}(t)\rangle_0
=\frac{2}{n}{\sum_k}'\frac{g\sin^2\omega_k}{\lambda_k}\label{alta}\\
&&\times\sin\lambda_kt[{\textstyle\frac{b-v\cos\omega_k}{\lambda_k}}
\sin\lambda_kt-i\cos\lambda_kt]\,,\nonumber\\
p_1(t)&=&\langle c^\dagger_j(t)c_j(t)c^\dagger_{j+1}(t)c_{j+1}(t)\rangle_0
\nonumber\\
&=&p^2(t)-\beta^2(t)+|\alpha^2(t)|\,,\label{p0}
\end{eqnarray}
where $j<n$ and in (\ref{p0}) we have applied Wick's theorem for vacuum
expectation values.

The corresponding results for $n=4,5$ and $14,15$ are also depicted in Figs.\
 \ref{f2}-\ref{f3}. It is seen that for low $g$, $C_2(t)$ presents sharp
type II resonances only below the outer peaks of $C_1$, and actually just below
the rightmost peak for small odd $n$. In order to understand this behavior, we
note that for $g\rightarrow 0$ and $b=b_k$,
\begin{equation}
\beta(t)\rightarrow{\textstyle\frac{2}{n}}\cos\omega_k\sin^2\lambda_kt\,,\;
|\alpha(t)|\rightarrow{\textstyle\frac{1}{n}}|\sin\omega_k\sin 2\lambda_kt|
 \,.\label{bag}\end{equation}
These limits can also be directly read from Eq.\ (\ref{apro}), as
$(2/n)\cos\omega_k$ is the average  $\langle
c^\dagger_jc_{j+1}\rangle=\sum_{k'}\cos\omega_{k'}\langle
c'^\dagger_{k'}c_{k'}\rangle/n$ in the state $|k,-k\rangle$ whereas $\alpha(t)$
is the average $\sum_{k'}\sin\omega_{k'}\langle
c'^\dagger_{k'}c'^\dagger_{-k'}\rangle/n$ in the full state (\ref{apro}). The
type II maxima of $C_{2}$ are then obtained for $\sin^2\lambda_kt=1$, leading
to
\begin{equation}
C^m_{2}\rightarrow {\textstyle\frac{4}{n}[|\cos\omega_k|-\sin\omega_k
\sqrt{1-\frac{4}{n}+\frac{4}{n^2}\sin^2\omega_k}}]\,,
 \label{cmn2}\end{equation}
in this limit at $b=b_k$. Eq.\ (\ref{cmn2}) is actually positive for
\[\sin^2\omega_k\leq {\textstyle[1-\frac{2}{n}+\sqrt{(1-\frac{2}{n})^2
+\frac{4}{n^2}}]^{-1}}\approx
  {\textstyle\frac{1}{2}+\frac{1}{n}+O(\frac{1}{n^2})}\,,\]
i.e., $\omega_k\alt\pi/4$ or $\omega_k\agt 3\pi/4$ ($|b_k|/v\agt 1/\sqrt{2}$)
for large $n$, so that they arise just beneath the outer peaks of $C_1$, the
strongest located at the rightmost peak for $n$ odd ($k=1/2$) and outermost
peaks for $n$ even ($k=1/2$ or $[n/2]-1/2$). Thus, type II resonances in
$C_{2}$ {\it remain also finite for $g\rightarrow 0$} but are of order
$n^{-1}$, becoming smaller than those of $C_1$ for large $n$
($C_2^m/C_1^m\propto \sqrt{2/n}$). The scaled concurrence $nC_2^m$ remains
nevertheless finite for large $n$.

For $n=3$ we {\it exactly} recover from (\ref{cmn2}) the previous result
$C^m_{2}=2/3$ for the type II peak. For $n=4$, Eq.\ (\ref{cmn2}) yields
$C^m_2=(2\sqrt{2}-1)/4\approx 0.46$, whereas for $n=5$ it leads to a single
peak at $\omega_k=\pi/5$, of height $\approx 0.41$. For $n=14$, there are sharp
type II peaks at the outer resonances, of height $\approx 0.22$, plus smaller
peaks at the next resonance, of height $\approx 0.08$, which rapidly fall below
the type I plateau. For  $n=15$ the visible type II peaks are asymmetric and
appear at $b_k/v\approx 0.98,0.81$ and $-0.91$, with heights $\approx
0.21,0.08$ and $0.15$.

For $g\rightarrow 0$ there are also type I maxima of $C_{2}$ at $b=b_k$,
visible in the central region (Fig.\ \ref{f3}). These maxima are broader and
occur at times determined by
\[\cos (2\lambda_kt)=
\frac{1-{\textstyle\frac{2}{n}}\sin^2\omega_k}{\sqrt{\sin^2\omega_k+
(1-{\textstyle\frac{2}{n}}\sin^2\omega_k)^2}}\,,\] (the first peak at
$t_1\approx\pi/(8\lambda_k)$ for $\omega_k\approx \pi/2$), where the
concurrence approaches for $g\rightarrow 0$ the value
 \begin{equation}
C^m_{2}\rightarrow {\textstyle\frac{2}{n}}
[\sqrt{(1-{\textstyle\frac{2}{n}}\sin^2\omega_k)^2+\sin^2\omega_k}-
(1-{\textstyle\frac{2}{n}}\sin^2\omega_k)]\,.
 \label{cmn1}\end{equation}
Since this is an increasing function of $|\sin\omega_k|$, i.e., a decreasing
function of $|b_k|$, the type $I$ maxima fall below those of type II for low
$|\sin\omega_k|$ ($|\sin\omega_k|\alt 0.66$ or $|b_k|/v\agt 0.75$ for large
$n$). Moreover, at the highest type I peak ($\omega_k\approx\pi/2$),
$C^m_{2}\approx 2(\sqrt{2}-1)/n$ for large $n$, which is just 21\% of the
highest type II peak ($C^m_{2}\approx 4/n$). For $n=3$ we also recover from
(\ref{cmn1}) the previous exact result $C^m_{2}=1/3$ in the type I plateau,
while for $n=2$ it yields the correct maximum value $C_2^m=1$. For $n=4$ and
$5$ we obtain $C^m_2\approx 0.14$ and $C^m_2\approx 0.07,0.2$ at the type I
peaks, while for $n=14,15$, $C^m_2\approx 0.07,0.06$ at the centermost type I
peak for $g\rightarrow 0$.

As $g$ increases, the lower type I resonances in $C_2^m$ become rapidly
smoothed out, merging into a broad plateau (Figs.\ \ref{f2},\ref{f3}).
Moreover, while for low $n$ the type II peaks remain visible even for large $g$
(Fig.\ \ref{f2}), as $n$ increases these peaks become as well superseded by the
type I plateau (Fig.\ \ref{f3}), which is discussed below. On the other hand,
for  $|b|\gg v,g$, we obtain, up to first order in $g/|b|$, $v/|b|$,
$C_2(t)\approx 2|\alpha(t)|\leq C_2^m$, with
\[C_2^m\approx \frac{4g}{n|b|}{\sum_k}'\sin^2\omega_k=\frac{g}{|b|}\,,\;\;\;
 n\geq 3\,,\]
in agreement with the previous result for $n=3$. In this limit, $C_2^m\approx
C_1^m/\sqrt{2}$.

\subsubsection{Temporal Evolution}
\begin{figure}[t]

 \centerline{\scalebox{0.7}{\includegraphics{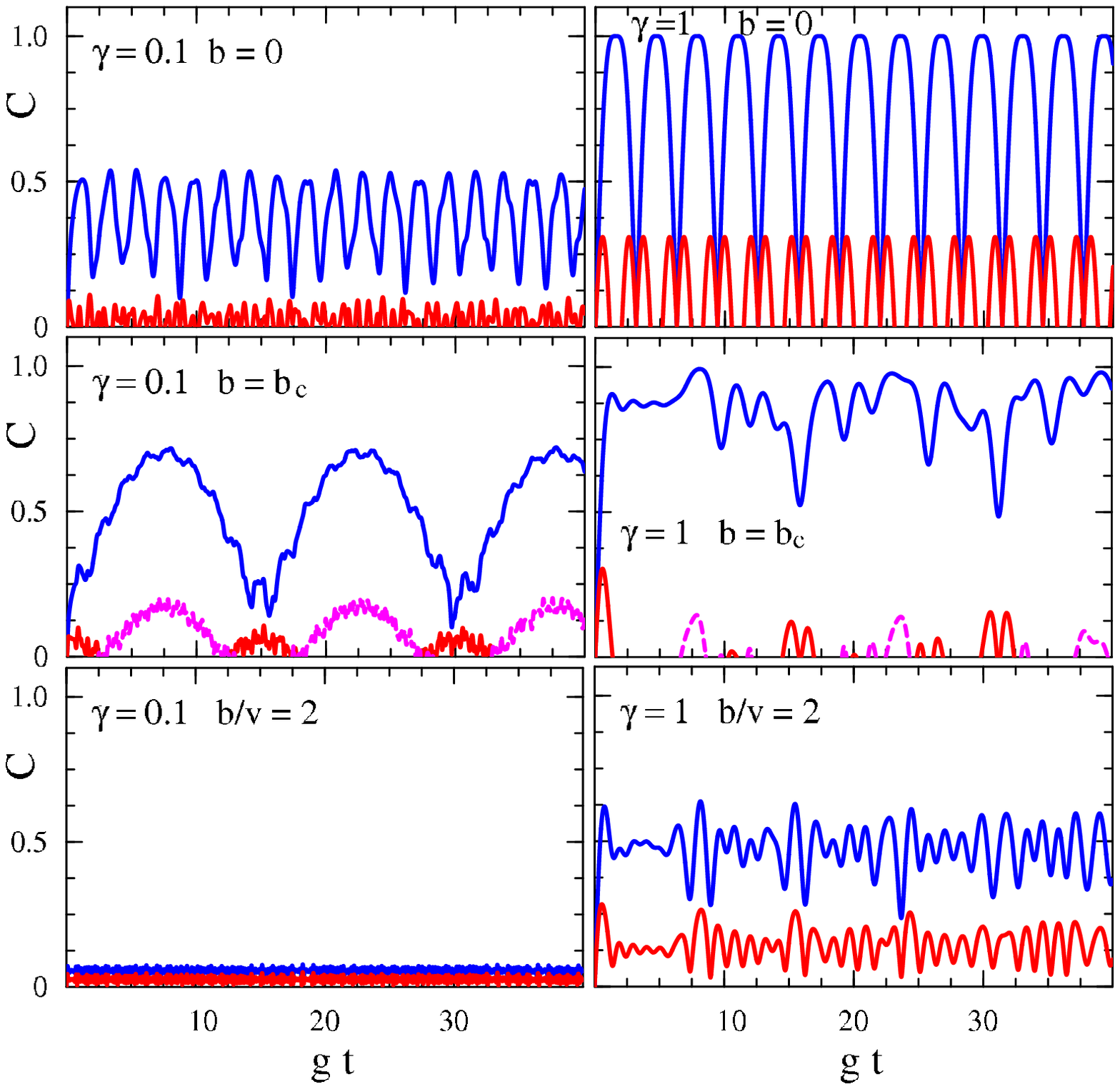}}}
\caption{(Color online). The evolution of $C_1(t)$ (upper curves in blue) and
$C_2(t)$ (lower curves, in red and pink) for $n=15$ at two different
anisotropies and different fields. The central panels depict the evolution at
the outer resonance $b_c/v=\cos(\pi/n)\approx 0.98$. Both the type I (red) and
type II (pink, dashed lines) sectors of $C_2(t)$  are indicated. }
 \label{f4}\vspace*{0.cm}
\end{figure}

Fig.\ \ref{f4} depicts $C_1(t)$ and $C_2(t)$ for $n=15$ at two different
anisotropies, {\it at} and {\it away} from resonances. For low $\gamma$ (left
panels), we observe a low frequency periodic-like evolution of $C_1(t)$ and
$C_2(t)$ at the outer resonance ($b/v\approx 0.98$), in agreement with
(\ref{ptg}) and (\ref{bag}), with $C_2(t)$ exhibiting regions of both type I
and type II entanglement, whereas for large fields $b=2v$ both $C_1(t)$ and
$C_2(t)$ become very small, with $C_2(t)$ of type I. Both $C_1(t)$ and $C_2(t)$
are also smaller for $b=0$ (with $C_2(t)$ again of type I), which here
corresponds approximately to a minimum of $C_1^m$ and $C_2^m$.

On the other hand, for $\gamma=1$ the emerging global entanglement is
non-negligible for all moderate fields, with saturation in $C_1$ reached for
$b\alt v$. In this case $C_2(t)$ does not follow the behavior of $C_1(t)$ for
low fields, where it strictly vanishes at finite time intervals, although for
large $b>v$ the evolution of $C_2(t)$ becomes again similar to that of $C_1(t)$
(with $C_2^m\approx C_1^m/\sqrt{2}$), and intervals of vanishing value are {\it
removed}. Thus, the {\it average} pairwise entanglement is in this case {\it
enhanced} by a large field $b\approx 2v$, in comparison with that for $b\approx
v$, as a consequence of the lower global entanglement. In other words, the
decoherence of the pair for large $\gamma$ due to the interaction with the spin
chain (representing here the environment for the pair) is prevented by large
fields.

It is also seen that the evolution for $\gamma=1$ ($g=v$) and $b=0$ is strictly
periodic. In this case $\lambda_k=v$ $\forall$ $k$ and Eqs.\ (\ref{alta})
become {\it independent} of $n$ for $n\geq 4$ and of the form
\begin{eqnarray}
p(t)&=&{\textstyle\frac{1}{2}}\sin^2 vt\,,\;\;\beta(t)=0,\;\;
\alpha(t)=-i{\textstyle\frac{1}{4}}\sin 2vt\,,\nonumber\\
C_1(t)&=&|\sin vt|\sqrt{2-\sin^2 vt}\,,\label{c1pe}\\
C_{2}(t)&=&|\sin vt|\,{\rm Max}\,[|\cos vt|-|\sin vt|/2,0]\,.\label{c2pe}
\end{eqnarray}
Hence, $C_1(t)$ reaches saturation when $|\sin vt|=1$,  whereas $C_{2}(t)$ has
maxima when $\cos 2vt=1/\sqrt{5}$, where $C_{2}(t)=(\sqrt{5}-1)/4\approx 0.31$,
and vanishes in the interval where $|\cos vt|<1/\sqrt{5}$ or when $\sin vt=0$.
The previous maximum of $C_2$ is already close to the maximum obtained for
large $\gamma$ (see below) and is higher than the resonant values for $n>9$.

 \begin{figure}[t]

 \centerline{\scalebox{0.5}{\includegraphics{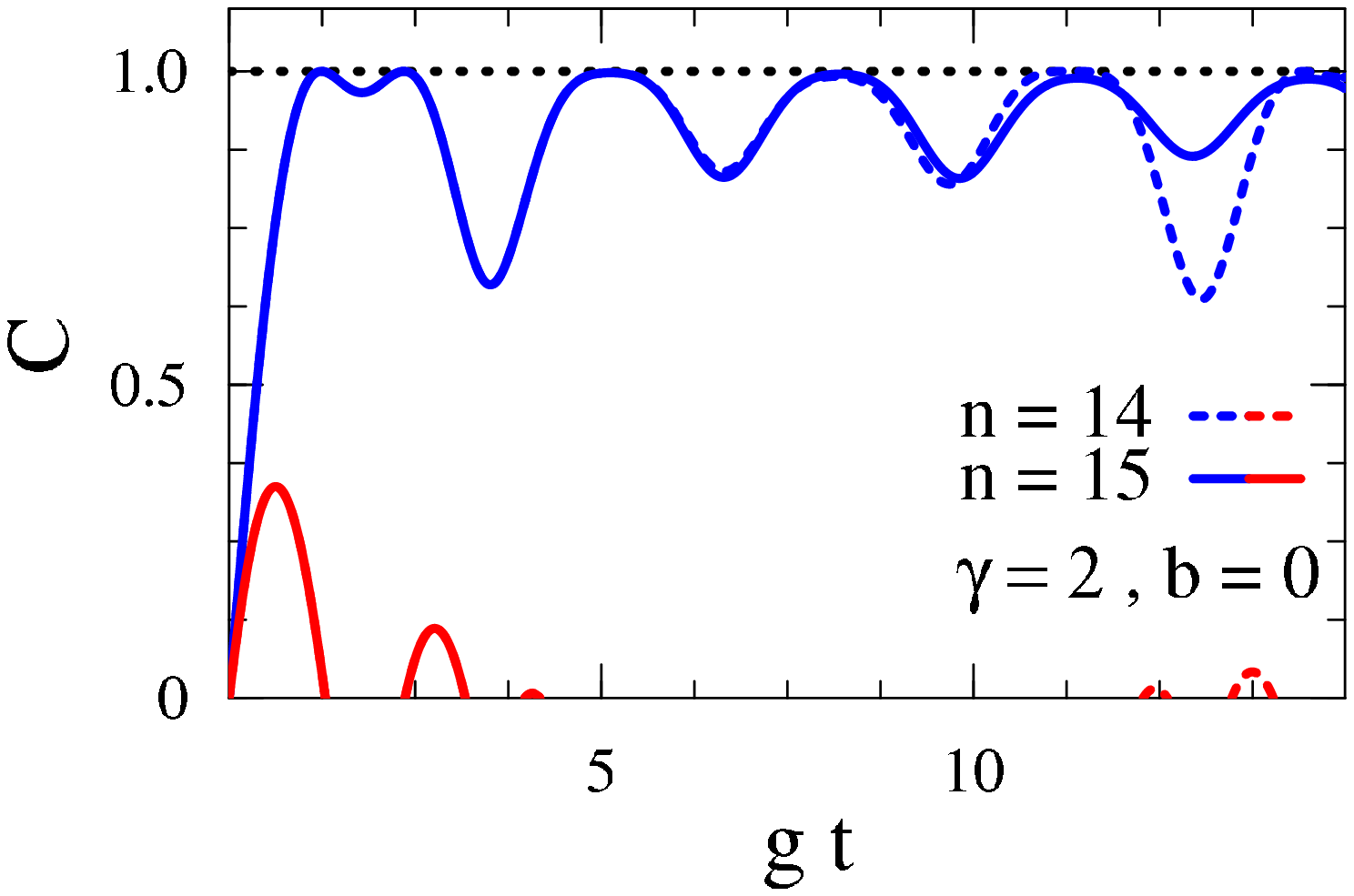}}}
\caption{(Color online). Evolution for large anisotropy and short times of
$C_1(t)$ (upper curves in blue) and $C_2(t)$ (lower curves, in red), for
neighboring odd-even systems.}
 \label{f5}
\end{figure}

Fig.\ (\ref{f5}) depicts the typical evolution for short times and large
anisotropy. As seen here, the plateau in the maximum concurrence $C_2^m$
arising for $g>(v,b)$ is originated  by the first maximum in the evolution of
$C_{2}(t)$, which exhibits in this region a prominent initial ``burst''
followed by intervals of vanishing value (i.e., decoherence of the pair) and
lower revivals (near the most prominent minima of $C_1(t)$). For $g\gg (b,v)$
and $n\agt 5$, the initial peak of $C_{2}$ occurs at $gt\approx 0.66$, with
height $C_{2}^m\approx 0.35$, and is practically {\it independent} of $n$. The
resonances in $C_2^m$, of order $n^{-1}$, become then rapidly covered by the
plateau as $n$ or $g$ increases. This initial peak can be approximately
reproduced by a fourth order expansion of $C_2(t)$, given for $n\geq 5$ by
\begin{eqnarray}
C_2(t)&\approx&{\textstyle gt[1-\frac{1}{2}gt-\frac{1}{6}t^2(v^2+b^2+3g^2)}
\nonumber\\&&+{\textstyle\frac{1}{12}}gt^3(2b^2+3g^2-v^2)\,.
 \nonumber\end{eqnarray}
Nonetheless, odd-even differences and $n$-dependence do arise for longer times
($gt\agt 10$ in the case of Fig.\ \ref{f5}) and affect the revivals of $C_2$.

Let us finally mention that as the resonances arising for low $\gamma$ develop
their first maximum at $t_k=\pi/(2g\sin\omega_k)$, the relevant timescale for
their observation is $\tau\approx\hbar/(\gamma v)\approx\tau_v/\gamma$, where
$\tau_v\approx\hbar/v$ is the operation time associated with the hopping
strength $v$, and should be smaller than the characteristic decoherence time
$\tau_d$ of the chain determined by its interaction with the environment. This
limits the smallness of the anisotropy (i.e., $\gamma\agt \tau_v/\tau_d$) and
hence the sharpness of the peaks. For instance, if $\gamma=0.1$ and $v\approx
0.02$ meV, which is a typical strength for realizations based on quantum dots
electron spins coupled through a cavity mode \cite{I.99}, $\tau\approx 3\times
10^{-10}s$, which is smaller than the typical decoherence time \cite{I.99}. On
the other hand, the results for $C_2$ represent the evolution of the
entanglement of an adjacent pair in the present spin chain environment, and
indicate that resonances remain finite at the pairwise level in such scenario.

\section{Conclusions}

We have examined the entangling capabilities of a finite anisotropic $XY$ chain
with constant parameters for an initially completely aligned state in the
transverse direction. The exact analytical results obtained (valid for all $n$)
show that the maximum attainable entanglement exhibits for low anisotropy
$\gamma$ a clear resonant behavior as a function of the transverse magnetic
field, with peaks at those fields where the effective quasiparticle energies
$\lambda_k$ are minimum and vanish for $\gamma=0$. At these fields, the energy
levels become then degenerate for $\gamma\rightarrow 0$ and the aligned state
remains mixed with its degenerate partner for arbitrarily small but non-zero
$\gamma$. The height of these resonances remains thus finite for
$\gamma\rightarrow 0$ and their width is proportional to the anisotropy,
implying a fine field sensitivity apt for efficient control, although the time
required to reach the peak is proportional to $\gamma^{-1}$ and the height
decreases as the number of qubits increases. The resonances are notorious in
the maximum global entanglement between one-qubit and the rest of the chain,
and are present as well in the entanglement of other global partitions.

They also arise in the maximum pairwise concurrence, and can be of both spin
parities, although they are of lower height and decrease more rapidly with $n$,
being hence more easily smoothed out for increasing $\gamma$.  Here we have
shown that type II (I) resonances become dominant at large (low) critical
fields for adjacent pairs, those of type II being extremely narrow. Another
feature is that odd-even differences in the resonant behavior remain
appreciable for moderate $n$, odd chains exhibiting field sign sensitivity both
in the global and pairwise peaks. On the other hand, saturation can be reached
in the global $E_1$ entanglement within a certain field window above a
threshold anisotropy ($\gamma\approx 1$ for large $n$), but not in the pairwise
entanglement, whose maximum exhibits instead a broad low plateau for large
$\gamma$ and hence low field sensitivity. Let us finally remark that resonances
of the present type will also occur for non-adjacent pairs as well as for other
geometries or interaction ranges, although details (i.e., relative widths and
strengths) may certainly differ from the present ones and are currently under
investigation.

RR acknowledges support of CIC of Argentina.


\begin{thebibliography}{999}

\bibitem{S.35}E.\ Schr\"odinger, Naturwissenschaften {\bf 23}, 807 (1935);
              Proc.\ Cam.\ Philos.\ Soc.\ {\bf 31}, 555 (1935).
\bibitem{Be.93}C.H.\ Bennett et al., Phys.\ Rev.\ Lett.\ {\bf 70}, 1895
    (1993); Phys.\ Rev.\ Lett.\ {\bf 76}, 722 (1996).
\bibitem{Ek.91}A.K.\ Ekert, Phys.\ Rev.\ Lett.\ {\bf 67}, 661 (1991);
               Nature {\bf 358}, 14 (1992).
\bibitem{Di.95}D.P.\ DiVincenzo, Science {\bf 270}, 255 (1995).
\bibitem{Be.00}C.H.\ Bennett and  D.P.\ DiVincenzo, Nature (London)
               {\bf 404}, 247 (2000).
\bibitem{NC.00}M.A.\ Nielsen and I. Chuang, {\it Quantum Computation and
               Quantum Information}, Cambridge Univ. Press (2000).
\bibitem{Be.96}C.H.\ Bennett, D.P.\ DiVincenzo, J.A.\ Smolin,
               W.K.\ Wootters, Phys.\ Rev.\ {\bf A 54}, 3824 (1996).
\bibitem{W.98}S.\ Hill and  W.K.\ Wootters, Phys.\ Rev.\ Lett.\ {\bf 78}, 5022
              (1997);W.K.\ Wootters, Phys.\ Rev.\ Lett.\ {\bf 80}, 2245 (1998).
\bibitem{ON.02} T.J.\ Osborne and M.A.\ Nielsen, Phys.\  Rev.\ {\bf A 66},
               032110 (2002).
\bibitem{OS.02}A.\ Osterloh et al, Nature (London) 416, 608 (2002).
\bibitem{V.03}G.\ Vidal, J.I. Latorre, E. Rico, A. Kitaev,
               Phys.\ Rev.\ Lett.\ {\bf 90}, 227902 (2003).
\bibitem{VW.04}F. Verstraete, M.A. Martin-Delgado, J.I. Cirac,
               Phys.\ Rev.\ Lett.\ {\bf 92}, 087201 (2004).
\bibitem{T.04} T. Roscilde, P. Verrucchi, A. Fubini, S. Haas, V. Tognetti,
              Phys.\ Rev.\ Lett.\ {\bf 93}, 167203 (2004).
\bibitem{LSM.61} E. Lieb, T. Schultz and D. Mattis, Ann. Phys.\ (NY) 16, 407
                 (1961).
\bibitem{S.99} A. Sachdev, {\it Quantum Phase Transitions},
               Cambridge Univ.\ Press (1999).
\bibitem{LDV.98} D.\ Loss and D.P.\ DiVincenzo,
                 Phys.\ Rev.\ {\bf A 57}, 120 (1998);
                 G.\ Burkard, D.\ Loss, and D.P.\ DiVincenzo,
                 Phys.\ Rev.\ {\bf B 59}, 2070 (1999).
\bibitem{I.99} A.\ Imamo\~glu, D.D.\ Awschalom, G.\ Burkard, D.P.\ DiVincenzo,
                 D.\ Loss, M.\ Sherwin, and A.\ Small, Phys.\ Rev.\ Lett.\
                 {\bf 83}, 4204 (1999).
\bibitem{Bos.04} S.C. Benjamin, S. Bose, Phys.\ Rev.\ Lett.\ {\bf 90} 247901
                (2003); Phys.\ Rev.\ {\bf A} 70, 032314 (2004).
\bibitem{L.02} J. Levy, Phys.\ Rev.\ Lett.\ {\bf 89}, 147902 (2002).
\bibitem{D.03} U. Dorner, P. Fedichev, D. Jaksch, M. Lewenstein, P. Zoller,
              Phys.\ Rev.\ Lett.\ {\bf 91}, 073601 (2003);
               J.J. Garc\'{\i}a-Ripoll, M.A. Martin-Delgado, J.I. Cirac,
                {\it ibid} {\bf 93}, 250405 (2004).
\bibitem{MSS.01} Y. Makhlin, G. Sch\"on and A. Shnirman, Rev. Mod. Phys.
              {\bf 73}, 357 (2001)
\bibitem{Ar.01}M.C.\ Arnesen, S. Bose and V. Vedral, Phys.\ Rev.\ Lett.\
               {\bf 87}, 017901 (2001).
\bibitem{Wi.02}X.\ Wang, Phys.\ Rev.\ {\bf A 66}, 044305 (2002);
               Phys.\ Rev.\ A {\bf 66}, 034302 (2002).
\bibitem{RC.05} R. Rossignoli, N. Canosa, Phys.\  Rev.\  {\bf A 72},
                012335 (2005); N. Canosa, R. Rossignoli, {\bf A 73} 022347
                (2006).
\bibitem{AOF.04} L. Amico, A. Osterloh, F. Plastina, R. Fazio, G.M. Palma,
                Phys.\ Rev.\ {\bf A 69} 022304 (2004).
\bibitem{SSL.04} A. Sen(De), U. Sen, M. Lewenstein, {\bf A 70}, 060304R (2004);
                 {\it ibid} {\bf A 72}, 052319 (2005).
\bibitem{HK.05} S.D.Hamieh, M.I.Katsnelson, Phys.\ Rev.\ {\bf A 72} 032316
                (2005).
\bibitem{HgK.06}Z.Huang, S. Kais, Phys.\ Rev.\ {\bf A 73}, 022339 (2006).
\bibitem{KRB.05}M. Koniorczyk, P. Rapcan, V. Buzek, Phys.\ Rev.\ {\bf A 72},
                022321 (2005).
\bibitem{CKW.00}V. Coffman, J.Kundu, W.K. Wootters, Phys.\ Rev.\ {\bf A 61},
                052306 (2000).
\bibitem{OV.06} T.J. Osborne, F. Verstraete, Phys.\ Rev.\ Lett.\ {\bf 96},
                220503 (2006).
\bibitem{DC.00} W.\ D\"ur and J.I.\ Cirac, Phys.\ Rev. {\bf A 61}, 042314
                (2000).
\bibitem{RS.80} P. Ring, P. Schuck, {\it The Nuclear Many-Body Problem},
                Springer, NY (1980).

\end{thebibliography}
\end{document}